\input phyzzx.tex
\tolerance=1000
\voffset=-0.0cm
\hoffset=0.7cm
\sequentialequations
\def\rl{\rightline}

\def\t1{{\tilde 1}}

\def\t{\theta}

\REF{\PRE}{P. Hayden and J. Preskill, JHEP {\bf 0709} (2007) 120, [arXiv:0708.4025].}
\REF{\SCR}{Y. Sekino and L. Susskind, JHEP {\bf 0810} (2008) 065, [arXiv:0808.2096].}
\REF{\COM}{L. Susskind, [arXiv:1402.5874]; [arXiv:1403.5695]; [arXiv:1411.0690].}
\REF{\WOR}{J. Maldacena and L. Susskind, Fortsch. Phys. {\bf 61} (2013) 781, [arXiv:1306.0533].}
\REF{\EDI}{E. Halyo, [arXiv:1502.01979]; [arXiv:1503.07808].}
\REF{\SHE}{S. H. Shenker and D. Stanford, JHEP {\bf 1403} (2014) 067, [arXiv:1306.0622]; JHEP {\bf 1412} (2014) 046, [arXiv:1312.3296].}
\REF{\SHO}{D. Stanford and L. Susskind, Phys. Rev. {\bf D90} (2014) 12, 126007, [arXiv:1406.2678]; D. A. Roberts, D. Stanford and L. Susskind, JHEP {\bf 1503} (2015) 051, [arXiv:1409.8180].}
\REF{\ADS}{J. Maldacena, Adv. Theor. Math. Phys. {\bf 2} (1998) 231, [arXiv:hep-th/9711200]; S. Gubser, I. Klebanov and A. Polyakov, Phys. Lett. {\bf B428} (1998) 105,
[arXiv:hep-th/9802109]; E. Witten, Adv. Theor. Math. Phys. {\bf 2} (1998) 253, [arXiv:hep-th/9802150].}
\REF{\LEN}{L. Susskind, [arXiv:hep-th/9309145].}
\REF{\EDIW}{E. Halyo. [arXiv:1403.2333]; [arXiv:1406.5763].}
\REF{\BUT}{L. Susskind, [arXiv:1311.7379].}



\singlespace
\rl{SU-ITP-15/10}
\pagenumber=0
\normalspace
\medskip
\bigskip
\titlestyle{\bf{Complexity Near Horizons}}
\smallskip
\author{ Edi Halyo{\footnote*{e--mail address: halyo@stanford.edu}}}
\smallskip
\centerline {Department of Physics} 
\centerline{Stanford University} 
\centerline {Stanford, CA 94305}
\smallskip
\vskip 2 cm
\titlestyle{\bf ABSTRACT}

We generalize the concept of complexity near horizons to all nondegenerate black holes. For Schwarzschild black holes, we show 
that Rindler observers see a complexity change of $S$ during proper time $1/\kappa$ which corresponds to the creation of a causal patch with proper length $1/\kappa$ inside the horizon. We attempt to describe complexity in the horizon CFT and the Euclidean picture.

\singlespace
\vskip 0.5cm
\endpage
\normalspace

\centerline{\bf 1. Introduction}
\medskip

There are good reasons to believe that black holes behave like quantum computers. For example, quantum circuits[\PRE] can reproduce
the black hole scrambling time[\SCR]. Therefore, the time evolution of black holes can be described as a quantum computation. If these ideas are correct, we should try to understand not only the thermodynamic properties of black holes such as temperature and entropy but also quantum computational properties such as complexity.

The relation between complexity and black holes was established in refs. [\COM] for $AdS$ black holes at the Hawking--Page transition. First, one can associate the coordinate distance to the horizon with complexity. The near horizon region consists of layers of increasing complexity as one approaches the horizon. Second, (the maximal analytical extension of) these $AdS$ black holes are ``two--sided". The two sides which can be considered to be two separate black holes are connected by an
Einstein--Rosen bridge (ERB) at a fixed time[\WOR]. The ERB has a length and volume that grow linearly with time which can be related to the complexity of the black hole. Finally, the ERB can be described as a collection of causal patches (CPs) that are continuously created inside the horizon. At any given time, the total entropy of the CPs created is equal to complexity.

In this paper, we generalize the relation between complexity, ${\cal C}$, and the coordinate distance to the horizon, $\ell$, to all nondegenerate black holes. The near horizon region of these black holes is Rindler space which consists of layers of different complexity as a function of $\ell$. The exact relation between ${\cal C}$ and $\ell$ is determined by the Rindler observer at a proper distance $1/\kappa$ to the horizon where $\kappa$ is the surface gravity. We show that Rindler observers see a change in complexity of $\Delta {\cal C}=S$ in a proper time interval of $1/ \kappa$. For Schwarzschild black holes in any dimension, this proper time
exactly corresponds to the proper length of a CP created inside the horizon which is also accompanied by the same change in complexity. Thus, even though the increasing ERB volume  and the creation of CPs inside the horizon are not observable, Rindler observers see the corresponding change in complexity.

We speculate that the complexity of the layers near the horizon is related to the complexity of the horizon CFT state that describes the black hole. These CFT states have $c/12=L_0=E_R$ where $E_R$ is the dimensionless Rindler energy[\EDI]. The increase in complexity towards the horizon may be described by the spread of nonzero coefficients in the CFT state wavefunction. We also relate complexity to
the Euclidean Rindler picture of black holes which gives us new insights into $\dot {\cal C}$ and entropy.

This paper is organized as follows. In the next section, we briefly review the results of refs. [\COM] on complexity that are
relevant for us. In section 3, we derive the relation between complexity and coordinate distance to the horizon for nonextreme black holes. For Schwarzschild black holes, we relate the change in complexity that is observed by a Rindler observer to the creation of CPs behind the horizon. In section 4, we describe how the change in complexity may be described in the horizon CFT.
Section 5 includes our conclusions and a discussion of our results.

\bigskip
\centerline{\bf 2. Complexity and Black Holes}
\medskip

Consider a black hole to be a quantum computer with $S$ qubits. Such a system reaches thermal equilibrium, i.e.
a uniform temperature, $T$, and maximum entropy of $S$ at the thermalization time $t_{therm} \sim S^p$. After $t_{therm}$, 
the system becomes time independent from a thermodynamic point of view.
However, from the perspective of quantum computation the system continues to evolve. This is best seen by looking at the computational
complexity of the system which continues to grow well beyond $t_{therm}$. In fact there are good reasons to believe that
complexity increases linearly with time, i.e. ${\cal C}=ST t$ since it is extensive and the only time scale for the thermal system in equilibrium is $1/T$[\COM]. The same result can also be obtained from the quantum circuit model of black holes[\PRE].
${\cal C}$ increases until the system reaches the maximum complexity ${\cal C}_{max}=e^S$ at time $t_{comp} \sim e^S$.

For black holes, it is not clear that complexity corresponds to a property that is measurable. However, in ref. [\COM], it was shown that, in the context of $AdS$ black holes at the Hawking--Page transition, complexity can be related to the coordinate distance to the horizon. A bulk degree of freedom at a coordinate distance $\ell$ from the horizon corresponds to a CFT operator on the boundary which is constructed from a boundary region of size $t$ given by
$$t=R_{AdS}log \left({R_{AdS} \over \ell}\right) \quad, \eqno(1)$$
where $R_{AdS}$ is the $AdS$ radius. Using the relation ${\cal C}=ST t$ we obtain the relation between the coordinate distance to horizon and complexity
{\footnote1{In ref. [\COM] the left--hand side of eq. (2) is squared which is due to the fact that eq. (1) holds up to numerical factors of $O(1)$.}}
$${\ell \over R_{AdS}}=e^{-{\cal C}/S} \quad. \eqno(2)$$
We conclude that the near horizon region of a black hole is made of layers of increasing complexity.
We stress that eq. (2) has been derived strictly for $AdS$ black holes with radius $R_{AdS}$. 
{\footnote2{Relations like eqs. (1) and (2) also hold for shock wave geometries[\SHE,\SHO].}}
In this paper, we will generalize it to all nonextreme horizons, i.e. those that reduce to Rindler space in the near horizon region.

On the other hand, the linearly increasing complexity of an $AdS$ black hole (with radius $R_{AdS}$) 
can be also related to the volume of the two--sided ERB and the total entropy of the CPs created beyond the horizon. An $AdS$ black hole is ``two--sided" and has a ERB that connects both sides at fixed time. The proper length
of the ERB increases linearly with time just like ${\cal C}$ does. As a result, the ERB volume, which is the length times the maximum cross sectional area, also grows linearly with time as[\COM]
$${dV \over dt}={{8 \pi G R_{AdS}} \over (D-1)} ST \quad, \eqno(3)$$
which leads to the relation between the volume of the ERB and complexity
$${\cal C}={(D-1) \over {GR_{AdS}}} V \quad. \eqno(4)$$
Unfortunately, it does not seem like the volume of the ERB is an observable quantity. The ERB can also be understood as a collection
of CPs, i.e. intersections of the past light cones of points on the singularity with the final slice inside the horizon[\COM]. 
The geometry of a CP is the product of $S^{D-2}$ and a finite interval.
The CPs are each created with coordinate length (i.e. the time interval inside the horizon) 
$$\Delta t=2 \int _0^{r_f} {dr \over {|f(r_f)|}} \quad, \eqno(5)$$
and have a covariant entropy of $S$. Above, $f(r_f)$ is the square of the redshift factor at the radius of the maximum ERB cross sectional area, $r_f$. The number CPs created in a given time $t$ is $N=t/\Delta t$. The total entropy of all the
CPs is the complexity of the black hole, i.e. $N S={\cal C}$. According to this picture, inside the horizon, there is a growing ERB with a length (or volume) proportional to time. Equivalently, the number of CPs created beyond the horizon, $N={\cal C}/S$ increases linearly with time.

\bigskip
\centerline{\bf 3. Complexity Near Nondegenerate Horizons}
\medskip

In this section, we generalize the relation between complexity and the coordinate distance to the horizon to all nondegenerate horizons. Consider any nonextreme black hole with a generic metric of the form
$$ ds^2=-f(r)~ dt^2+ f(r)^{-1} dr^2+ r^2 d \Omega^2_{D-2} \quad, \eqno(6)$$
in D-dimensions. The horizon is at $r_h$ which is determined by
$f(r_h)=0$. If in addition, $f^{\prime}(r_h) \not =0$, the near horizon geometry is described by Rindler space. Near the
horizon, we have $r=r_h +y$ where $y<<r_h$, which leads to the near horizon metric
$$ds^2=-f^{\prime}(r_h)y~ dt^2+(f^{\prime}(r_h)y)^{-1} dy^2+ r_h^2 d \Omega^2_{D-2} \quad. \eqno(7)$$
In terms of the proper radial distance, $\rho$, obtained from $d\rho=dy/\sqrt{f^{\prime}(r_h)y}$ the metric becomes
$$ds^2=-\kappa^2 \rho^2 dt^2+d \rho^2+ r_h^2 d \Omega^2_{D-2} \quad, \eqno(8)$$
where the surface gravity is $\kappa=f^{\prime}(r_h)/2$.
A freely falling (inertial) observer sees a flat metric in the coordinates $(T,X)$ which are given by
$$T=\rho cosh(\kappa t) \qquad  X=\rho sinh(\kappa t) \quad. \eqno(9)$$
This shows that the Rindler observer at fixed $\rho$ has an acceleration of $\kappa$ relative to the
inertial observer. The Rindler observer can be at any proper distance to the horizon. However, we will take her to be at $\rho=1/\kappa$ since only at this point the proper time is equal to the coordinate time $t$. This Rindler observer is special
since complexity is proportional to $t$ which is exactly her proper time.

Now, the coordinate distance of the Rindler observer to the horizon in the $(T,X)$ coordinates is
$$\ell=X-T={e^{-\kappa t} \over \kappa} \quad. \eqno(10)$$
Thus, in the flat coordinate frame, the Rindler observer (at $\rho=1/\kappa$) is getting exponentially closer to the horizon due to her acceleration. Inverting eq. (10) we get
$$t=-{1 \over \kappa} log (\kappa \ell) \quad, \eqno(11)$$
which is the analog of eq. (1). We now assume ${\cal C}=S\kappa t$ which is a factor $2\pi$ larger than what was assumed for the $AdS$ case in ref. [\COM] since $T=\kappa/2 \pi$. We can change the definition of ${\cal C}$ up to a constant since the although 
$\dot {\cal C}=ST$ on dimensional grounds
there is no way to determine the exact constant of proportionality from first principles. Then, the relation between $\ell$ and ${\cal C}$ becomes
$$\kappa \ell=e^{-{\cal C}/S} \quad. \eqno(12)$$
This is the analog of eq. (2) for all nondegenerate horizons. For $AdS$ black holes with a horizon radius of $R_{AdS}$, 
$\kappa=1/R_{AdS}$ and we get eq. (2) as required. The near horizon region has the expected layered structure in which a complexity is associated with every coordinate distance $\ell$ to the horizon. ${\cal C}$ increases as we approach the horizon and diverges there.
The complexity defined by eq. (12) is based on the assumption that $\rho=1/\kappa$. However, the Rindler observer can be at any $\rho$. For $\rho \not=1/\kappa$ we get
$$\kappa \ell=\kappa \rho e^{-\kappa t}=e^{-{\cal C}/S} \quad. \eqno(13)$$
For $\rho<1/\kappa$ eq. (13) generalizes eq. (12). Note that in this case the initial complexity is positive.
On the other hand, for $\rho>1/\kappa$ the initial complexity is negative. This hard to interpret so when
eq. (13) leads to ${\cal C} <0$ we will set ${\cal C}=0$. However, in this case,
since complexity increases with time it turns positive at $t_0=(1/\kappa) log(\kappa \rho)$. Therefore, for $t>t_0$, $C>0$ and eq. (13) gives a meaningful definition complexity. 

It is interesting to note that the problem of negative complexity does not arise in de Sitter space. Consider the static patch for de Sitter space which has a metric of the form in eq. (6) with 
$$f(r)=\left(1-{r^2 \over R^2}\right) \quad, \eqno(14)$$
where the de Sitter radius, $R$, is the reciprocal of the Hubble constant. It is easy to see that near the de Sitter horizon the metric reduces to that of Rindler space with $\kappa=1/R$. Then, we can define complexity of a point located at $r$ as
$${{R-r} \over R}=e^{-{\cal C}/S} \quad. \eqno(15)$$
In this case, since $0 \leq r \leq R$ negative complexites do not arise.

Consider now a Rindler observer at $\rho=1/\kappa$. Since we can write eq. (12) also as
$${\cal C}=-S log(\kappa \ell) \quad, \eqno(16)$$
in a time interval of $\Delta t=1/\kappa$, (say from $t=n/\kappa$ to
$t=(n+1)/\kappa$) she moves closer to the horizon from a layer with ${\cal C}=nS$ to one with ${\cal C}=(n+1)S$ giving rise to the change $\Delta {\cal C}=S$. Thus, during this interval, the Rindler observer is exposed to a change in complexity that is equal to the black hole entropy. As we will show below, for Schwarzschild black holes, this exactly corresponds to the creation of a CP behind the horizon. The result $\Delta {\cal C}=S$ is independent of $\rho$; i.e. it holds not only for the special Rindler observer at $\rho=1/\kappa$ but for all Rindler observers. Consider one at $\rho < 1/\kappa$ for whom ${\cal C}$ is defined by eq. (13). Then
solving for ${\cal C}$ we obtain the general relation
$${\cal C}=S \kappa t-Slog(\kappa \rho) \quad. \eqno(17)$$
Since $\rho$ is fixed for a given observer, the second term above is constant and for a time interval of $1/\kappa$ we find again $\Delta {\cal C}=S$. However, for these observers the proper time is $t_{prop}=\rho t$ and not equal to the coordinate time.
We note that the relation $\dot {\cal C}=ST$ fixes ${\cal C}$ only up to a constant which is usually set to zero. However, for Rindler observers at $\rho<1/\kappa$ the constant does not vanish and is given by eq. (17). On the other hand, for $\rho>1/\kappa$ the constant term in eq. (17) is negative which leads to ${\cal C}<0$ as we saw above.

We now relate the complexity change seen by Rindler observers to the creation of CPs behind the horizon.
For simplicity, we consider only $D$--dimensional eternal Schwarzschild black holes for the CP calculations. The maximal analytical extensions of these black holes are  ``two--sided" and have an ERB that connects both sides, i.e. the two black holes, at a fixed time. For these black holes the metric is given by eq. (6) with
$$f(r)=\left(1-{\mu \over r^{D-3}}\right) \quad, \eqno(18)$$
where $\mu=16 \pi G_D M/(D-2)A_{D-2}$ and $A_{D-2}$ is the volume of the $(D-2)$--dimensional unit sphere. The cross sectional area of the ERB is found by maximizing the function[\COM]
$$g(r)=|f(r)|r^{2D-4}=\mu r^{D-1}-r^{2D-4} \quad, \eqno(19)$$
with respect to $r$ which gives the radius, $r_f$, at which the ERB cross section is maximal
$$r_f=\left({{D-1} \over {2D-4}}\right)^{1/{D-3}} \mu^{1/{D-3}} \quad. \eqno(20)$$
The surface gravity of the black hole is given by
$$\kappa={f^{\prime}(r) \over 2}={{D-3} \over {2 \mu^{1/{D-3}}}} \quad. \eqno(21)$$

As mentioned above, behind the horizon, CPs with a cylindrical geometry of $S^{D-2}$ times an interval are generated.
The coordinate length of the interval is 
$$\Delta t=2 \int _0^{r_f} {dr \over {|f(r_f)|}}={2r_s \over {(D-3)\sqrt{|f(r_f)|}}} \quad, \eqno(22)$$
where $f(r_f)=\sqrt{|(3-D)|/(D-1)}$ and $r_s=1/\mu^{1/(D-3)}$ is the Schwarzschild radius. 
{\footnote3{Eq. (22) is correct up to a dimension dependent numerical factor that is of $O(1)$ for $4\leq D \leq 10$. We neglect this factor in the following since all the formulas involving ${\cal C}$ are correct up to numerical factors.}}

The generation of a CP increases the complexity by
$\Delta {\cal C}=S$ which exactly corresponds to the increase of complexity observed by a Rindler observer during a time interval of $1/\kappa$. By using eqs. (20) and (22), we can write the coordinate length of a CP as 
$$\Delta t={1 \over {\kappa \sqrt{|f(r_f)|}}} \quad. \eqno(23)$$

We now relate the proper time interval of the Rindler observer during which she sees $\Delta {\cal C}=S$ to the proper length of the CP inside the horizon. 
(We note that inside the horizon it is the $t$ direction in the metric in eq. (6) that is space--like and gives a length.)
The proper length of the CP is 
$$L=\sqrt{|f(r_f)|} \Delta t={1 \over {\kappa}} \quad. \eqno(24)$$
This is exactly the proper time interval of the Rindler observer at $\rho=1/\kappa$ which is equal the coordinate time interval, $1/\kappa$. 

The picture that emerges is as follows. For a proper time interval of $1/\kappa$, a Rindler observer at $\rho=1/\kappa$ sees a
change in complexity of $\Delta {\cal C}=S$. This proper time interval in the near horizon region
corresponds to the proper length of the CP created inside the horizon. The creation of a CP inside the horizon also leads
to the same change in complexity.
Every time the Rindler observer sees a change of $S$ in complexity, a CP is created inside the horizon. During this interval, the 
volume of the ERB increases by
$$\Delta V={{S G } \over {\kappa (D-3)}} \quad. \eqno(25)$$

Even though we have explicitly considered only $D$--dimensional Schwarzschild black holes, our definition of complexity near
nondegenerate horizons should apply to all nonextreme black holes with $\kappa \not= 0$. However, for nonextreme black holes with an extremal limit, such as charged or rotating black holes, it is not straightforward to define ERBs and CPs. In addition, in de Sitter space, the space behind the horizon is expands exponentially making it hard to define final slices and CPs even though
we can define the complexity near the horizon by eq. (15). Thus, in all these cases, even though the definition of ${\cal C}$ is outside the horizon is known, its relation to the physics behind the horizon is not clear.
{\footnote4{I would like to thank Lenny Susskind for raising these issues.}}

\bigskip
\centerline{\bf 4. Complexity in the Horizon CFT}
\medskip

As a result of the AdS/CFT correspondence[\ADS], for AdS black holes, the complexity associated with a given coordinate distance to the horizon refers to the complexity of the boundary CFT state (or the operator that creates that state). Obviously, we do not have such a holographic relation in flat backgrounds. Nevertheless, it is tempting to speculate that, in these cases complexity
may be related to the state of horizon CFTs that live in the very near horizon region. 

Recently, it has been shown that nonextreme black holes can be described by states of a $D=2$ CFT with $c/12=L_0=E_R$ where $E_R$ is the dimensionless Rindler energy[\EDI].
These states have a degeneracy of $exp(2\pi E_R)$ that leads to the black hole entropy of $S=2\pi E_R$[\LEN,\EDIW]. Let us denote the degenerate CFT states by $|i>$ where $i=1, \ldots,exp(2\pi E_R)$.
These microstates have the same dimensionless energy, $E_R$ so their energy measured by a Rindler observer at $\rho=1/\kappa$ is
$E_i=E_R/\rho=\kappa E_R$.
Now in order to describe a chaotic system we assume that the energy levels are slightly nondegenerate by exponentially small amounts of $\Delta E_i \sim e^{-S} \kappa E_R$[\BUT]. The important point is that different linear combinations of the $|i>$ may have different complexities.
We would like to associate the complexity of a layer near the Rindler horizon with the complexity of a horizon CFT state. 

Eq. (12) tells us that at a distance $\ell=1/\kappa$ from the horizon ${\cal C}=1$. This corresponds to a simple CFT state given by 
$$|\phi>=\sum_i^{e^{2\pi E_i}} |i> \quad, \eqno(26)$$
where all the coefficients are equal to $1$, i.e. the spread in the coefficients is zero. Clearly, this is a very special state. 
At closer coordinate distances, the CFT state $|\phi>$ is given by a linear combination of $|i>$ but now the coefficients are spread by some nonzero amount.
At exponentially small coordinate distances that correspond to the maximal complexity ${\cal C}_{max} \sim e^S$, the coefficients are
separated by $O(1)$. Thus, from the CFT point of view, the spread of coefficients in the CFT state corresponds to increasing complexity. As a result, we may define the complexity of a CFT state as $e^S$ times the generic phase difference between any two coefficients.

Since, for a Rindler observer,  the coordinate distance to the horizon is related to time $t$, we should be able to see the spreading of coefficients in the time evolution of the CFT state. At $t=0$, the observer is at $\ell=1/\kappa$ with 
${\cal C}=1$ which is described by the simple initial state $|\phi(0)>$ in eq. (26). 
At time $t$ the CFT state becomes
$$|\phi(t)>=\sum_i e^{-i E_i t} |i> \quad, \eqno(27)$$
where the different energies, $E_i$, are separated by $\sim e^{-S} \kappa E_R$. After a time interval of $t=1/\kappa$ the phase
differences between different $|i>$ become around $\Delta E_i/\kappa \sim e^{-S} S$. Therefore, the complexity of this state is $S$.
This is what we found for the change in complexity experienced by a Rindler observer in the time interval of $1/\kappa$. 
At any time $t$, the generic phase difference between the coefficients is $\Delta E_i t \sim e^{-S} S \kappa t$ so that 
${\cal C}=S \kappa t$ which is exactly the assumed form of ${\cal C}$ at any given time. At the exponentially large time $t_{comp}=e^S/\kappa$, the phase differences generically become
$O(1)$ which leads to the maximal complexity $e^S$ for the CFT state.

Obviously, the above description of complexity is not rigorous. First of all,
we know nothing about the horizon CFT beyond $c$ and $L_0$. Second, we do not know the exact definition of computational
complexity for CFT states. Nevertheless, it seems that it is not too difficult to come up with a definition
of complexity for CFTs that behaves correctly.
 
Finally we would like to relate our results to the effective description of nondegenerate black holes in Euclidean Rindler space.
Defining dimensionless Euclidean Rindler time by $\tau=i \kappa t$, the metric in eq. (6) becomes
$$ds^2=\rho^2 d \tau^2+ d \rho^2+r_h^2 d\Omega^2_{D-2} \quad ,\eqno(28)$$
which looks like flat space in polar coordinates. The compact Euclidean time direction has period $2 \pi=1/T_R$ which is the inverse dimensionless Rindler temperature. The Euclidean coordinates are given by (where $\kappa=1$)
$$T=\rho cos \tau \qquad X=\rho sin \tau \quad. \eqno(29)$$
We see that the Euclidean dimensionless time becomes and angle and the
time evolution of a Rindler observer at fixed $\rho$ is just a rotation. The (Wald) entropy of the black hole is given by
$S=2 \pi E_R$[\LEN,\EDIW] where the dimensionless Rindler energy $E_R$ is canonically conjugate to $\tau$. It is also a holographic 
quantity given by the surface Hamiltonian of General Relativity over the horizon. In this space,
the rate of change in complexity is given by 
$${d{\cal C} \over  d\tau}=ST_R={S \over {2 \pi}}=E_R \quad. \eqno(30)$$
Thus, $E_R$ not only determines the black hole entropy but also has an information theoretic meaning as the rate of change in complexity. If the time evolution of the black hole can be described as a parallel quantum computation with two qubit gates, then $E_R$ is the number of computations per unit time.  

We saw that black hole entropy is $S=2 \pi E_R=\Delta \tau E_R$ where $\Delta \tau=2 \pi$ corresponds to one cycle around the Euclidean time direction.
This raises two questions: 1) why does one cycle give the black hole entropy? and 2) what is the meaning of
all the other cycles? The answers are clear. In Rindler space, ${\cal C}=E_R \tau$ and its integral over one cycle corresponds
to $\Delta {\cal C}$ over the time it takes to create a CP behind the horizon, i.e. $\Delta {\cal C}=S$. We can say that black hole entropy is given by $S=2 \pi E_R$ because we are actually computing the complexity change in the dimensionless time interval 
$2 \pi$ that happens to be the entropy.
If we think about the black hole time evolution as a quantum computation, it is clear that each cycle 
corresponds to a parallel computation that takes time $1/T_R=2 \pi$ and involves all degrees of freedom to give the correct entropy.

If the above picture is basically correct it raises another question. The relation $S=2 \pi E_R$ is exact; in fact in gives the Wald entropy of any nonextreme black hole in any theory of gravity[\EDIW]. On the other hand, the equations that involve complexity are valid only up to numerical factors. It is then surprising that complexity arguments can count black hole entropy exactly.

\bigskip
\centerline{\bf 5. Conclusions and Discussion}
\medskip

In this paper, we generalized the notion of computational complexity to all nondegenerate horizons. We found that (accelerating)
Rindler observers at fixed proper distance to the horizon, $\rho$, are exposed to layers of increasing complexity as a function of time. Such observers see a change of complexity of $\Delta {\cal C}=S$ in a time interval of $1/\kappa$. For Schwarzschild black holes, this exactly matches the creation of a CP behind the horizon that has the proper length $1/\kappa$ and leads to the same change in ${\cal C}$. It is also accompanied by an increase in the length and volume of the ERB inside the horizon. We also speculated that complexity can be associated with that of a horizon CFT state that describes the black hole.

We saw that even though we can define complexity near all nondegenerate horizons, for charged and rotating black holes it is hard to
associate this complexity with ERBs and CPs inside the horizon. Clearly, it is important to see whether this is possible and if it is whether the definition of complexity inside the horizon matches the one outside it given by eq. (12). de Sitter space suffers from a similar problem due to the exponentially increasing volume outside the de Sitter horizon. It would be interesting to see if a definition of complexity behind the de Sitter  horizon that matches eq. (15) can be found.


\bigskip
\centerline{\bf Acknowledgments}

I would like to thank Lenny Susskind for useful discussions and the Stanford Institute for Theoretical Physics for hospitality.

\vfill

\refout

\end
\bye